\documentclass[12pt]{article}
\usepackage[polish,english]{babel}
\usepackage[latin1]{inputenc}
\usepackage{times}
\usepackage[T1]{fontenc}
\usepackage{epsfig}
\newcommand{\p}{\partial}
\newcommand{\be}{\begin{equation}}
\newcommand{\ee}{\end{equation}}
\begin{document}

\title{Self-similarity for V-shaped field potentials - further examples }

\author{H. Arod\'z, $\;\;$ P. Klimas $\;$ and $\;$ T. Tyranowski \\$\;\;$ \\  Institute of Physics,
Jagiellonian University, \\ Reymonta 4, 30-059 Cracow, Poland}

\date{$\;$}

\maketitle

\begin{abstract}
Three new models with V-shaped field potentials $U$ are considered: a complex scalar field $X$ in 1+1 dimensions
with $U(X) = |X|$, a real scalar field $\Phi$ in 2+1 dimensions with $U(\Phi) = | \Phi|$, and a real scalar
field $ \varphi$ in 1+1 dimensions with  $U(\varphi) = \varphi \: \Theta(\varphi)$ where $\Theta$ is the step
function.  Several explicit, self-similar solutions are found. They describe interesting dynamical processes,
for example, `freezing' a string in a static configuration.

\end{abstract}

\vspace*{2cm} \noindent PACS: 05.45.-a, 03.50.Kk, 11.10.Lm \\
\noindent Preprint TPJU - 4/2007

\pagebreak

\section{ Introduction}

Lagrangian for a field, let it be a real scalar field $\phi$, usually contains a kinetic part, and a potential
$U(\phi)$ which, in the case of physical models, is assumed to be at least bounded from below. In most models,
however, $U(\phi)$ is a smooth function of $\phi$ with  isolated absolute minima which are reached at finite
values of $\phi$. Furthermore, if $\phi_0$ is one of such minima, the second derivative $ U''(\phi_0) = m_0^2
\geq 0 $ exists and plays the role of length scale ($ \lambda_0 \sim 1/m_0$) in the model. For a weak classical
field such a model is reduced to a free field model, and when quantizing it one can apply the harmonic
oscillator paradigm with particle creation and annihilation operators. Models with these features are very
popular -- justly so because they have plenty of important applications.

Said above notwithstanding, one can find interesting models which do not fit that description. In particular,
there exist physically well-motivated models such that $U''$ is infinite at the minimum of $U$ \cite{1}.
Specifically, the field potential $U$ is V-shaped around the minimum, hence the first derivative has a
discontinuity at $\phi = \phi_0.$ In papers \cite{1, 2, 3} we have investigated certain relatively simple models
of this kind with a single real scalar field in 1+1 dimensions and, moreover, with the field potential invariant
under the transformation $ \phi(x)  \rightarrow - \phi(x)$. Among the most interesting findings was a scaling
symmetry of the `on shell' type. This symmetry is universal in the sense that it does not depend on neither the
number of space dimensions nor the number of fields.

In the present paper we continue the investigations of  field theoretic models of that kind.  Our goal here is
to find self-similar solutions in  models which are less restricted than the ones considered in the previous
papers: we allow for more fields, or more space dimensions, or smaller symmetry.  We consider three new models.
In the first one we have a complex scalar field in 1+1 dimensions (Section 2), the second involves a real scalar
field in 2+1 dimensions (Section 3), and in the third model the field potential is not invariant under the
transformation $ \phi(x) \rightarrow - \phi(x)$ and has degenerate minimum extending from $ \phi=0$ to $-
\infty$ (Section 4). These models are interesting already on purely theoretical ground as examples of V-shaped
field potentials and this is our main motivation. Nevertheless, these models can have applications. For example,
the first two can be regarded as models of pinning of a string or a membrane, respectively, and the third model
describes the process of depinning of a planar string. It should be stressed that total energy is conserved in
these models - the pinning occurs because of the dynamics and not because of a dissipative loss of energy.

In the case of V-shaped potentials the pertinent Euler-Lagrange equations are nonlinear in a rather unusual way
- as a rule they contain a discontinuous function of the field. Restricting the considerations to the sector of
self-similar fields is a natural simplifying step which reduces by one the number of independent variables.
Theory and applications of self-similar solutions of nonlinear evolution equations is a well-established,
important branch of theory of nonlinear systems, see, e.g., \cite{4, 5}. It turns out that in the  sector of
self-similar fields there is quite interesting dynamics which can be seen, e.g.,  from the explicit solutions
which are presented in subsequent Sections.

\section{Complex scalar field in 1+1 dimensions}

The string in three dimensional space is attracted to a straight line, which we call the $z$-axis. The two
directions perpendicular to it are denoted as the $X_1, X_2 $ directions. For notational simplicity, we use
dimensionless coordinates $z, X_1, X_2$ - the physical coordinates are obtained from them by multiplying by a
unit of length. Position of the string at a fixed time $t$ is given by two functions $X_1(z,t), X_2(z,t)$ of $z
\in (-\infty, \infty),$ which give Cartesian coordinates of points of the string in the $(X_1, X_2)$ planes
\footnote{Also $t$ is a dimensionless variable proportional to the physical time.}  . We will consider only the
cases of a stretched string, so that $X_1, X_2$ are single-valued functions of $z$.

We are interested in the dynamics of this system in the special case when the attractive force has the potential
\begin{equation}
U(X_1, X_2) = |X|,
\end{equation}
where $X = X_1 + i X_2$ is the complex number representation of points from the planes perpendicular to the
$z$-axis. Then, the evolution equation for the string, in the approximation discussed in the previous Section,
has the form \be (\p_t^2 - \p_z^2) X_k = - \frac{X_k}{|X|}, \ee where $k=1,2.$ Thus, the attractive force has
constant modulus equal to +1. Equation (2) can be written as  \be (\p_t^2 - \p_z^2) X = F(X), \ee where \be F(X)
= - \frac{X}{|X|}. \ee Here we have assumed that $X \neq 0$. However, one should take into account the fact that
$X=0$ is physically acceptable configuration of the string - the string just rests on the $z$-axis. In order to
formally  include $X=0$ to the set of solutions of the evolution equation we assume that $F(X) =0$ if $X=0$. Of
course $F(X) $ is discontinuous at $X=0$.

It should be noted that we are looking for so called weak solutions of Eq. (2), \cite{6, 7}. When such a
solution is inserted in Eq. (2) one should obtain an identity only when the both sides are integrated with
arbitrary test function of the variables $t, z, $ while in the case of strong solutions the identity is obtained
immediately after the substitution. Actually, the weak solutions are the right ones in the context of
Euler-Lagrange equations because the stationary action principle has precisely the weak form
\[
\int dt dx \: \delta X_i \frac{\delta S}{\delta X_i} = 0,
\]
where $S$ is the action functional and $\delta X_i$ arbitrary test functions.

The potential $U$ is V-shaped: plot of $U(X_1, X_2)$ has the form of a symmetric cone with the tip at the point
$X_1 = X_2 =0$. Equation (3) possesses the scale invariance: if $X$ is a solution of it,  then \be
X_{\lambda}(z,t) = \lambda^2 \: X(\frac{z}{\lambda}, \frac{t}{\lambda}) \ee is a solution too for any $\lambda >
0$ \footnote{Transformations with $\lambda <0$ can be obtained by combining transformations (5) with the
space-time reflection $z \rightarrow -z, \; t \rightarrow -t$. }. By definition, the self-similar solutions are
invariant with respect to these transformations.

It is convenient to use a polar Ansatz for $X:$
\[
X = \rho \: \exp(i \chi),
\]
where $\rho$ and $\chi$ are real.  Then
\[
\frac{X}{|X|} = sign(\rho) \exp(i \chi),
\]
where the signum function takes values $\pm 1,$ or $0$ if $\rho=0$.  Eq. (3) is equivalent to  the following two
equations \be \rho \left(\p_t^2 - \p_z^2\right)\: \chi + 2 \left(\p_t\chi \: \p_t\rho - \p_z\chi \:\p_z
\rho\right) =0, \ee \be \left(\p_t^2 - \p_z^2\right) \rho - \left(\p_t\chi\: \p_t\chi - \p_z\chi\: \p_z
\chi\right)\: \rho= - sign(\rho). \ee In the case of constant phase $\chi$ this set of equations reduces to the
signum-Gordon equation considered in \cite{3}. For this reason, we focus here on self-similar solutions of Eqs.
(6, 7) with non constant $\chi$.

Self-similar Ansatz for $X$ has the form \be \chi = H(y), \;\rho = z^2 S(y), \; y = \frac{t}{z}.     \ee Then
Eq. (6, 7) are reduced to the set of nonlinear ordinary differential equations for $S$ and $H$: \be (1-y^2) (S
H'' + 2 H' S') + 2y S H' =0, \ee \be (1-y^2) ( S'' - H'^2 S) + 2 y S' - 2 S = - sign( S), \ee where $'$ stands
for the derivative $d/dy$.  Notice that the presence of the factor $1-y^2$ implies that values of $ H'$ and $
S'$ may have finite jumps at $ y = \pm 1$. One way to see it is as follows: the finite jump of the first
derivatives would lead to Dirac's delta terms $\sim \delta( y \pm 1)$ in the second derivatives, but such terms
are not harmful because $(1-y^2) \: \delta(y \pm 1) =0.$

Finding solutions with non constant $\chi$ is greatly simplified by the observation that  Eq. (9) can be written
in the following form
\[
\left( \frac{S^2 H'}{1-y^2} \right)' =0.
\]
Hence, \be  S^2 H' = c_0 (1-y^2), \ee with $c_0$ being a real constant. Therefore, we can eliminate $H'$ from
Eq. (10),  \be (1-y^2) S'' - \frac{c_0^2 (1-y^2)^3}{S^3} + 2 y S' - 2 S = - sign(S). \ee

Equation (12) has the following solution, valid in the region $|y| <1,$ \be S_1 = \alpha (1-y^2), \ee where
$\alpha$ is a real constant related to $c_0$: \be 4 \alpha^4 + c_0^2 = |\alpha|^3. \ee Inserting formula (13) in
Eq. (12) and integrating the resulting equation for $H$ we find that \be H(y) =  \frac{ d_0}{2}
\ln\frac{1+y}{1-y} + h_0, \ee where $h_0$ is a constant and $d_0 = c_0/\alpha^2.$ Formula (14) implies that \[
|\alpha| = \frac{1}{4 + d_0^2}.\] Therefore, \be S_1 = \pm \frac{1-y^2}{4+d_0^2}. \ee Let us stress that
solution (13) is valid only in the interval $ |y| <1.$ If we take $|y|  \geq 1, $ then instead of (14) we obtain
the condition \[ 4 \alpha^4 + c_0^2 = -   |\alpha|^3,\] which implies that $\alpha = 0 = c_0.$  Therefore, the
Ansatz (13) considered in the region $ |y| \geq 1$ yields only the trivial solution $S_0=0$. The partial
solutions $S_0, \; S_1$  match each other at $y = \pm 1.$ Together they form the continuous solution on the
whole $y$-axis. The first derivative with respect to $y$ has a finite jump at $y= \pm 1$, but this is allowed
for by Eq. (10).

Inserting $z^2 S_1$ and $H$ in the polar Ansatz we obtain the final form of the self-similar solution for $t
\geq 0$ (at a given time $t
>0$ the solution $S_1$  covers the $ z
>t$ and $ x < -t $ parts of the $z$ axis, while  the interval $ |z| \leq t$ is covered by the trivial solution $S_0
=0$):
\begin{equation}
X(z,t) = \left\{ \begin{array}{ccc} \frac{1}{4 + d_0^2}\;(z^2 - t^2)\; \exp\left(i \frac{d_0}{2}
\ln\frac{z+t}{z-t} + i h_0 \right) & \;\;\;\; \mbox{if} \;\;\; & |z| \geq t,  \\
0 & \;\;\;\; \mbox{if} \;\;\; & |z| \leq t.
\end{array}\right.
\end{equation}
The constant phase $h_0$ is not interesting - it just corresponds to global rotations around the $z$-axis. The
minus sign present in formula (16) has been included in this phase. It is clear that the distance between the
string and the $z$-axis, equal to $|X|$, quadratically grows from 0 at the points $z=\pm t$ to $\infty$ when $z
\rightarrow \pm \infty.$ Much more interesting is the behaviour of the phase of $X$ at a fixed time $t$. It
describes the winding of the string around the $z$-axis. At  $z \rightarrow \infty$ the phase $\chi$ is equal to
$h_0$. When $z$ decreases, the phase increases by $2 \pi$  with each step from $z$ to $z - \Delta z$, where
\[
\Delta z = \frac{z^2  -t^2}{z + \coth(2\pi/d_0)t}
\]
(we have assumed here that $d_0 > 0)$. Thus, the string winds around the $z$-axis infinitely many times as $z
\rightarrow t$. Similar behaviour of the phase is found when $ z \rightarrow -t$,  the only difference is that
the string winds in the opposite direction.

\begin{center}
\begin{figure}[tph!]
\hspace*{1cm}
\includegraphics[height=8cm, width=9.5cm]{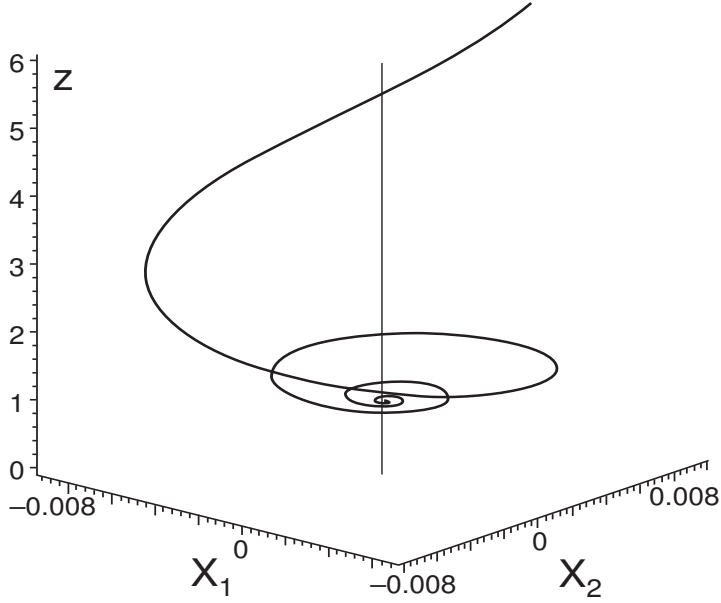}
\caption{ The $z \geq t$ branch of  self-similar solution (17).  }
\end{figure}
\end{center}

Such unusual behaviour of the string might create a suspicion that it is unphysical for the following reason:
that there is an infinite amount of energy accumulated in the finite region around the point $z=t$ (and,
symmetrically, around $z = -t$). It turns out that this is not the case. The total energy $E_T$ of the piece of
the string between the points with the coordinates $z=t$ and $z = t +T$, where $T > 0$, is given by the formula
\[ E_T = \int^{t+T}_t dz \; \left( \frac{1}{2} \p_tX_i \p_tX_i +  \frac{1}{2} \p_z X_i \p_zX_i + |X| \right).
\]
Elementary calculation shows that
\[
E_T = \frac{1}{2} \frac{T}{4 + d_0^2}\: (T + t) \:(T + 2t).
\]
Furthermore,  the length of that piece of the string is also finite:
\[  L_T(t) = \int^{t+T}_t dz \; \sqrt{|\p_z X|^2 +1}
\approx \frac{t T}{\sqrt{4+d_0^2}}.
\]
The last formula is valid for $ t \gg T$ and $ t \gg \sqrt{4 + d_0^2}$.

Let us notice that if we put $c_0 =0$, then $d_0 =0,$ and  the solution has the simple form
\[ X = \left\{ \begin{array}{ccc} \frac{1}{4} (z^2 - t^2) \exp(ih_0) & \;\;\; \mbox{for} \;\;\; & |z| \geq t  \\
0 & \;\;\; \mbox{for} \;\;\; & |z| leq t. \end{array} \right.
\]
This particular solution coincides with one of self-similar solutions of signum-Gordon equation considered in
\cite{3}. In this case the string lies in one plane containing the $z$-axis. Solutions with $c_0 \neq 0$ are
more general - the string winds around the $z$-axis. In the case of $c_0 =0$ Eq. (12) has the same form as in
the signum-Gordon model \cite{3}. Thus, the `centrifugal' term $ \sim c_0^2$ in Eq. (12) is the only effect of
presence of the second scalar field $X_2$, as far as the self-similar solutions are concerned.

Solution (17) can be regarded as the solution of an initial value problem for the string being pinned to the
$z$-axis. The initial data  are given at the time $t=0$ which corresponds to $y=0$. Formula (17) implies that
\be X(z, t=0) = \frac{1}{4 + d_0^2} \exp(i h_0) z^2, \;\; (\p_t X)(z, t=0) =  \frac{d_0 i}{d_0^2+4} \exp(i h_0)
z. \ee Thus, at the initial moment all points of the string lie in one plane (the one with azimuthal angle
$h_0$), and the initial velocities $(\p_t \vec{X})(z, t=0) $ of the points of the string are perpendicular to
their initial position vectors $ \vec{X}(z, t=0) = (X_1, X_2)(z, t=0)$.

Let us end this Section with the observation that one can also obtain asymmetric solutions, i.e., such that $X
=0$ for $z <t$ or for $z > -t$. The corresponding initial data are obtained by multiplying both formulas (18) by
$\Theta(z)$ or $\Theta(-z)$. The point is that for $t>0$  $y=0$ corresponds to both $z = \infty$ and $z =
-\infty.$ Therefore, Eq. (12) should actually be considered separately in the regions $y>0$ and $y<0$, and there
is no justification for a requirement of continuity of $S$ at $y=0$!. For this reason we may take the trivial
solution $S_0$ in the whole region $y<0$ and $S_1$ in the interval $[0,1]$, or $S_0$ for all $y >0$ and $S_1$
for $y \in [-1, 0]$. These choices give the asymmetric solutions.  On the other hand, notice that $y = \infty$
and $y = - \infty$ correspond to $z=0$. Therefore, the behaviour of $S(y)$ at $y \rightarrow \infty$ is
correlated with the behaviour at $y \rightarrow - \infty$ because the function $X$ is  continuous at $z=0$.
However, this is not such a severe condition as it might seem because formula (8) for the $\rho$ function
contains  the factor $z^2$ which vanishes for $z=0$ - it relates only the terms $ \sim y^2$ in the function
$S(y)$ in the regions $ y > 1$ and $ y < -1$.

\section{ Real scalar field in 2+1 dimensions}

Let us consider a membrane (in three dimensional space) which is attracted to a plane.  The plane is
parametrised by the Cartesian coordinates $x, y$, and the elevation of the membrane at the moment $t$ over the
point $(x,y)$ of the plane  is denoted by $\Phi(x, y, t)$. We consider the membrane without overhangs, hence
$\Phi$ is a single valued, real function of $x, y$. The evolution equation has the form \be (\p_t^2 - \triangle)
\Phi = - sign(\Phi), \ee where $\triangle$ denotes the two-dimensional Laplacian. The corresponding field
theoretic potential again has the form (1), that is $U(\Phi) = |\Phi|.$ The scaling transformations which are
the symmetry of Eq. (19)  have the form
\[
\Phi_{\lambda}(x, y, t) = \lambda^2 \: \Phi(\frac{x}{\lambda}, \frac{y}{\lambda}, \frac{t}{\lambda}), \] where
$\lambda >0$.

For simplicity, we  consider only the axially symmetric configurations, hence $\Phi$ does not depend on the
azimuthal angle. In this case the evolution equation has the form \be (\p_t^2 - \p_r^2 - \p_r/r) \Phi = -
sign(\Phi), \ee where $r = \sqrt{x^2 + y^2}.$ The self-similar Ansatz \be \Phi = r^2 P(w), \;\; w = \frac{t}{r}
\ee reduces Eq. (20 ) to ordinary differential equation \be (1-w^2) P'' + 3 w P' - 4 P = - sign(P), \ee where
$'$ denotes the derivative $d / dw.$ The corresponding homogeneous equation \be (1-w^2) P'' + 3 w P' - 4 P = 0.
\ee  has the following linearly independent solutions \footnote{$P_1$ has been obtained by inserting a
polynomial Ansatz in Eq. (22), and $P_2$ from $P_1$ by the standard method \cite{8}}: \be P_1 = 1 + 2 w^2, \ee
\be P_{2} = \left\{
\begin{array}{l} (1 + 2 w^2) \arccos w - 3 w \sqrt{1-w^2} \;\;\; \mbox{if} \;\;\; |w| \leq 1, \\ -(1 + 2 w^2)
\ln|w - \sqrt{w^2-1}| - 3 w \sqrt{w^2 -1} \;\;\; \mbox{if} \;\;\; |w|
> 1. \end{array} \right. \ee
 From
them we can construct partial solutions of Eq. (22) \be P_+ = \alpha_1 P_1 + \alpha_2 P_2 + \frac{1}{4} \;\;
\mbox{if} \;\;\; P_+ > 0, \;\; P_- =  \beta_1 P_1 + \beta_2 P_2 - \frac{1}{4} \;\; \mbox{if} \;\;\; P_- < 0, \ee
where $ \alpha_{i},  \beta_i$ are constants. In general, these solutions are valid only in certain finite
intervals of the $w$ axis, determined by the inequalities $P_+
> 0, P_- < 0$ - for this reason we have called them the partial solutions.    There also exists the trivial
solution $  P=0$ (remember that $sign(0) =0$), and the static solutions \be P = \pm \frac{1}{4}. \ee

By glueing together these solutions one can produce various self-similar solutions of Eq. (22).  We will present
here just one class of such solutions, which describes how the membrane `freezes' in the static position. These
solutions involve the static solution and the partial solution $P_+$ which are glued together on the
`light-cone' $ r=t$, i.e., at the point $w=1$, see Fig. 2.  In accordance with Eq. (23), in this case it is
sufficient to demand only continuity of $P(w)$. This condition has the form $P_+(w=1) = 1/4$. Simple calculation
shows that it is equivalent to the condition $\alpha_1 =0$. Therefore, the corresponding solution $\Phi(r,t)$ of
Eq. (20) has the following form \be \Phi(r, t) = \left\{
\begin{array}{l} \frac{r^2}{4} \;\;\; \mbox{if} \;\;\;  r \leq t, \\
\alpha_2 \left[ (r^2 + 2  t^2) \arccos\frac{t}{r} - 3 t \sqrt{r^2 - t^2}\right] + \frac{r^2}{4} \;\;\;\;
\mbox{if} \;\;\;\; r \geq t.  \end{array} \right. \ee Here $t \geq 0$, and we have chosen that branch of
$\arccos w$ for which $\arccos 1 =0.$ Moreover, \be \alpha_2
> - \frac{1}{2 \pi}. \ee This last condition ensures that $ P_+ > 0$ on the whole half-line $ r \geq t. $ Thus,
with time larger and larger part of the membrane becomes static. It follows from the explicit form (28) of the
solution that at the initial time $t=0$ \be \Phi(r, t=0) =  \frac{1}{2} \left( \pi \alpha_2 + \frac{1}{2}
\right) r^2, \;\;\; \left.\p_t \Phi\right|_{t=0} = - 4 \alpha_2 r.  \ee If, for example, the initial slope of
the parabola is smaller than 1/4 ($\alpha_2 <0$), the initial velocity is positive, i.e., the parabola moves
upwards towards the static shape. Notice, however, that this is done in a manner consistent with the causal
dynamics implied by wave equation (20) - the freezing occurs at the  front $r=t$ which travels with the finite
velocity equal to +1.

\begin{center}
\begin{figure}[tph!]
\hspace*{1cm}
\includegraphics[height=7cm, width=11cm]{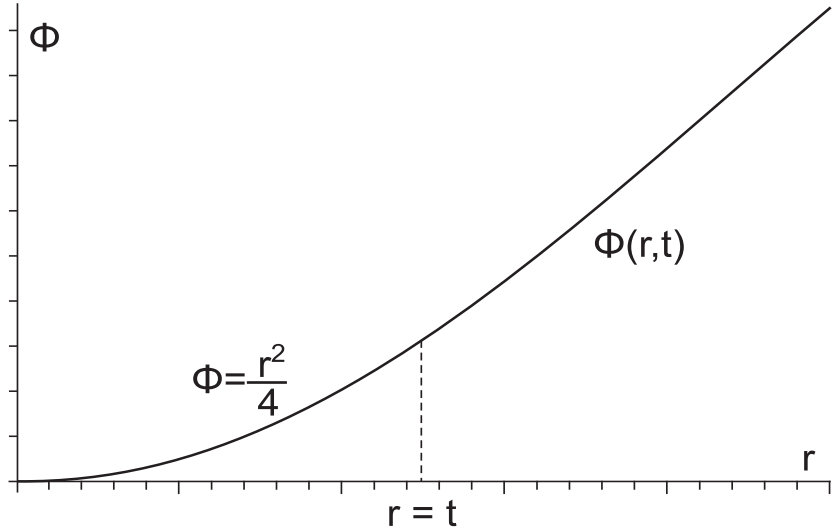}
\caption{ Snapshot  of  self-similar solution (28).  }
\end{figure}
\end{center}

One may expect that there are self-similar solutions of Eq. (19) which depend on the azimuthal angle $ \theta$
in the $(x, y)$ plane. The angle $\theta$ is invariant under the rescaling $ (x, y) \rightarrow (x/\lambda,
y/\lambda)$. The Ansatz for self-similar solutions could be taken in the form $ \Phi = r^2 P(w, \theta)$. We
have not investigated such solutions of Eq. (19).

\section{  $\Theta$-Gordon model in 1+1 dimensions}

As pointed out in \cite{1}, the scaling symmetry  (4) is shared by all (1+1)-dimensional models with V-shaped
field potential
\[
U(\varphi) = c_1 \: \varphi \: \Theta(\varphi)  - c_2  \: \varphi \: \Theta(-\varphi),
\]
where  $c_1, c_2$ are arbitrary real constants,  $\Theta$ denotes the step function \footnote{$\Theta(\varphi) =
1$ when $ \varphi >0$ and $\Theta(\varphi) = 0$ when $ \varphi \leq 0$.}, and $\varphi(z,t)$ is a real scalar
field. The corresponding wave equations have the form \be (\p_t^2 - \p_z^2) \varphi = - c_1 \Theta(\varphi) +
c_2 \Theta(-\varphi). \ee We are interested in the models such that for $ \varphi =0$  the potential energy has
minimum at $\varphi =0$ . For this, we have to assume that $ c_1 \geq 0, \; c_2 \geq 0.$ In the particular case
$c_1 = c_2 =1$ the potential $U$ is symmetric with respect to the reflection $ \varphi \rightarrow - \varphi$,
and Eq. (31) becomes the signum-Gordon equation considered in \cite{3}. Now we would like to investigate the
self-similar solutions in the case $c_1 =1, \:c_2=0,$ when Eq. (31) has the form \be (\p_t^2 - \p_z^2) \varphi =
- \Theta(\varphi). \ee For the obvious reason we will call this equation the $\Theta$-Gordon equation.

Physical context for considering equation of the form (32) is the depinning phenomenon.  This equation can be
regarded as describing the dynamics of a string in a plane which would be permanently pinned to the $z$ line
were it not for a constant bias force, which exactly compensates the pinning force on one side of the $z$ line
($ \varphi$ is just the deviation of the string from this line). Hence, the $\Theta$-Gordon equation describes
the dynamics of the string exactly at the depinning transition.

The Anstaz for self-similar solutions has the form \[ \varphi(z,t) = z^2 R(y), \] where $y = t/z$. Inserting it
in Eq. (32) we obtain the following equation
\[
(1-y^2) R'' + 2 y R' - 2 R = - \Theta(R) \Theta(z^2)
\]
where $'$ denotes the derivative $ d/ dy$.  This equation is to be satisfied in the weak sense, hence we may
drop the factor $\Theta(z^2)$ which differs from 1 only at the single point $z=0$. Therefore, $R(y)$ obeys
(again in the weak sense) the following equation \be (1-y^2) R'' + 2 y R' - 2 R = - \Theta(R). \ee Equation (33)
has the trivial solutions $ R =0, \; R = 1/2,$ which correspond, respectively, to $ \varphi =0$ and to the
static solution \be \varphi_s = z^2/2. \ee Furthermore, there are  the partial solutions \be R_+ = a_1 + a_2 y +
(a_1 -\frac{1}{2}) y^2 \;\; \mbox{if} \;\;\; R_+
> 0, \;\; R_- = b_1 (1 + y^2)  + b_2 y \;\; \mbox{if} \;\;\; R_- < 0, \ee where $ a_{i},
b_i$ are constants. As in the previous Sections, the partial solutions are valid only in certain intervals of
the $y$ axis, determined by the inequalities $R_+ > 0, R_- < 0$.

Glueing together the partial solutions one can obtain whole variety of self-similar solutions of Eq. (32)
\cite{9}. We present here certain interesting examples. \\
1.  Freezing in the static configuration. \\
This solution is depicted in Fig. 3.
\begin{center}
\begin{figure}[tph!]
\hspace*{1cm}
\includegraphics[height=6cm, width=10cm]{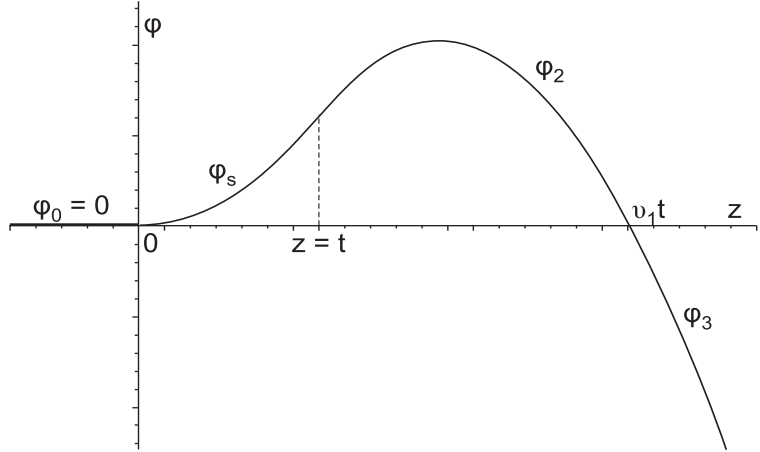}
\caption{ Snapshot of  self-similar solution (36).  }
\end{figure}
\end{center}
It has the following explicit form
\be \varphi(z,t) = \left\{ \begin{array}{lr}     0 &  z \leq 0,  \\
\frac{1}{2} z^2  & z \in[0,t], \\  \frac{1}{2(v_1 v_2 -1)} \left( v_1 t -z \right)
\left( z - v_2 t \right) & z \in [t, v_1 t], \\
S_0 z^2 + S_0' t z + S_0 t^2 & z \geq v_1 t,
\end{array}\right.
\ee where \[ v_1 = \frac{ S_0' + \sqrt{S_0^{'2} - 4 S_0^2}}{2 |S_0|}
> 1, \;\;\;v_2 = \frac{v_1}{2 v_1 -1} <1,
\]
and $ S_0 <0$. The parameters   $S_0^{'}, S_0$  are not independent -- they have to obey the equation \be
S_0^{'3} + (2 S_0 -1) S_0^{'2} - 8 S_0^3 + 4 S_0^2 - S_0 - 4 S_0^2  S_0^{'} =0,   \ee which follows from the
condition of continuity of the derivative $\p_z\varphi$ at the point $ z= v_1 t$. It turns out that  cubic
equation (37) has three real solutions when $S_0 <0$. The relevant root  $S_0'$ obeys the inequality $ S_0'
> 1/2 - 2 S_0$.

Initial configuration for solution (36) has the form
\[
\varphi = S_0 z^2 \Theta(z), \;\;\; \p_t\varphi = S_0' z \Theta(z).
\]

\noindent 2. Depinning of the string from the z-axis. \\
The function $ \varphi$ has the  shape presented in Fig. 4.
\begin{center}
\begin{figure}[tph!]
\hspace*{1cm}
\includegraphics[height=6cm, width=10cm]{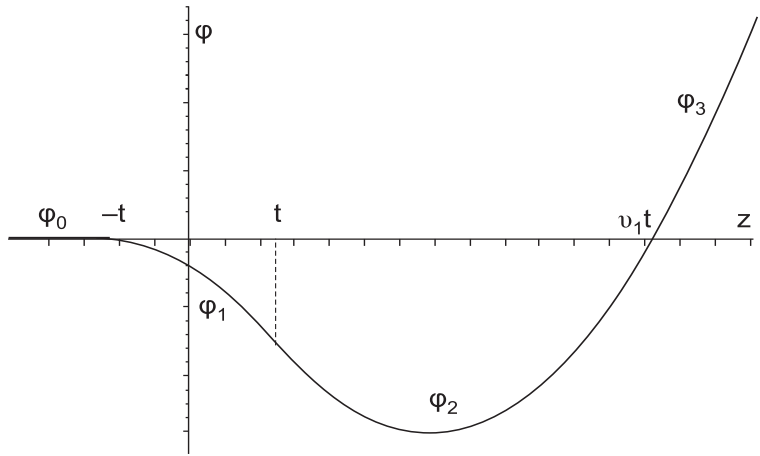}
\caption{ Snapshot of  self-similar solution (38).  }
\end{figure}
\end{center}
In this case
\be \varphi(z,t) = \left\{ \begin{array}{lr}     0 &  z \leq -t,  \\
- \alpha (z +t)^2  & z \in[-t, t], \\ - \beta  \left( v_1 t -z \right)
\left( z - \frac{1}{v_1} t \right) & z \in [t, v_1 t], \\
S_0 z^2 + S_0' t z + (S_0 - \frac{1}{2}) t^2 & z \geq v_1 t,
\end{array}\right.
\ee where $ S_0 >0$, and  \[ v_1 = \frac{ - S_0' + \sqrt{S_0^{'2} - 4 S_0^2 + 2 S_0}}{2 S_0}
> 1, \]
\[ \alpha =\frac{1}{4} ( 2 S_0 v_1 + S_0')  \frac{v_1 -1}{ v_1 +1} <1, \;\;\;\;  \beta = \frac{4 \alpha v_1}{(v_1
-1)^2}. \]

Solution (38) is valid provided that $v_1 >1$. This condition is satisfied if \be
 0 < S_0, \;\;\; S_0' < 1/2 - 2 S_0. \ee

\section{Remarks}

\noindent 1. We have presented examples of self-similar solutions which describe rather interesting dynamical
processes like the freezing in the static configurations or the depinning.   However, it is obvious that the
three models considered here have other self-similar solutions as well. In the case of $\Theta$-Gordon model one
can provide the complete list of such solutions \cite{9} analogous to the one given in \cite{3} for
signum-Gordon model. In the other two models such a complete list is probably out of our reach because of the
$S^{-3}$ term in Eq. (12), or possible azimuthal angle dependence in the case of membrane model. \\

\noindent 2. The kinetic part of evolution equation (3) contains  d'Alembert operator $ \p_t^2 - \p_z^2$, and
this is  standard for bosonic field. On the other hand, when we apply this equation in order to describe
evolution of the string we automatically make certain assumptions about the kind of string we consider. The
precise statement is that it is the string which has evolution equation of the form (3) when its world-sheet is
parametrized by the coordinates $t, z$. Of course, we would like to see a connection with Nambu-Goto string,
which has evolution equation of the form \be \frac{\p}{\p u^a} \left( \sqrt{-g} g^{ab} \frac{\p x^{\mu}}{\p u^b}
\right) =0, \ee where $ u^0 =t, u^3 =z, (x^{\mu}) = (t, X_1(z,t), X_2(z,t), z)$, $ \; (g^{ab}) $ is the inverse
to $(g_{ab}),$ and
\[
g_{ab} = \frac{\p x^{\mu}}{\p u^a} \frac{\p x^{\nu}}{\p u^b} \eta_{\mu\nu}.
\]
Let us take a slightly more general than  Minkowskian space-time metric $\eta_{\mu\nu}$: \be (\eta_{\mu\nu}) =
\mbox{diag}(1, - l_0^2, -l_0^2, -1), \ee where $l_0^2$ is a constant (equal to 1 in the  case of Minkowski
space-time). Simple calculations show that Nambu-Goto equation (40) is reduced to the equation
\[
\p_t^2 X_i - \p_z^2 X_i =0
\]
when \be l_0 \p_tX_i \ll 1, \;\;\; l_0 \p_z X_i \ll 1. \ee Interesting possibility to satisfy the conditions
(42) is to take a very small $l_0$. This would correspond to the Nambu-Goto string (or a linear defect like a
vortex) in an anisotropic medium which would effectively provide  metric (41). The term $F(X)$ in Eq. (3)
represents the pinning force with which the $z$ line attracts the string.

Analogous remarks can be made about the membrane discussed in Section 3 and about the planar string of Section 4. \\

\noindent 3.  The V-shaped field potentials $U(\phi)$ we consider should not be identified as potentials which
just contain the modulus of the pertinent field. Examples of such potentials can be found in \cite{10, 11}. The
fundamental difference is that in our models the second derivative $U''$ does not exist right at the minimum of
the potential, hence there is no preferred finite length (or mass) scale. Potentials considered in \cite{10, 11}
have the $\Lambda$ shape. In these cases $U''$ does not exist at a local maximum while at minima it exists --
that has much smaller impact on properties of the fields. \\

\end{document}